\documentclass{aa} 
\usepackage[varg]{txfonts}
\usepackage{natbib}
\usepackage{graphicx}
\usepackage[colorlinks=true,linkcolor=black,anchorcolor=black,citecolor=blue,filecolor=black,menucolor=black,runcolor=black,urlcolor=black]{hyperref}
\usepackage{graphicx}
\usepackage{subcaption}
\usepackage{float}
\usepackage{url}
\usepackage{arydshln}
 
\begin{document}

\title{Spot evolution on LQ~Hya during 2006--2017: temperature maps based on SOFIN and FIES data}
 
\titlerunning{Spot evolution on LQ~Hya during 2006--2017 \thanks{Based on observations made with the Nordic Optical Telescope,
    operated by the Nordic Optical Telescope Scientific Association at
    the Observatorio del Roque de los Muchachos, La Palma, Spain, of
    the Instituto de Astrofisica de Canarias.}} 
\authorrunning{Cole-Kodikara et al.}
  
  \author{Elizabeth M. Cole-Kodikara\inst{1}
  \and Maarit J. K\"apyl\"a\inst{2,3}
    \and Jyri J. Lehtinen\inst{2,3}
  \and Thomas Hackman \inst{4}
  \and Ilya V. Ilyin \inst{1}
  \and Nikolai Piskunov \inst{5}
  \and Oleg Kochukhov \inst{5}}      

\institute{
Leibniz-Institute for Astrophysics Potsdam, An der Sternwarte 16, 14482 Potsdam, Germany
\and
Max Planck Institute for Solar System Research, Justus-von-Liebig-Weg, 3, G\"ottingen, Germany
  \and ReSoLVE Centre of Excellence, Department of Computer Science, Aalto University, Finland
  \and Department of Physics, P.O. Box 64, FI-00014 University of Helsinki, Finland
  \and Department of Physics and Astronomy, Uppsala University, Box 516, 751 20 Uppsala, Sweden}
 
\abstract{LQ~Hya is one of the most frequently studied young solar analogue stars.
Recently, it has been observed to show intriguing behaviour by analysing long-term 
photometry:
during 2003--2009, a coherent spot structure migrating in the rotational frame
has been reported by various authors, but since that time 
the star has entered a chaotic state where coherent structures seem to have disappeared 
and rapid phase jumps of the photometric minima occur irregularly over time.}
{LQ~Hya is one of the stars included in the SOFIN/FIES long-term monitoring 
campaign extending over 25 years. 
Here we publish new temperature maps for the star during 2006--2017, 
covering the chaotic state of the star.}
{We use a Doppler imaging technique to derive surface 
temperature maps from high-resolution spectra.}
{From the mean temperatures of the Doppler maps we see a weak but 
systematic increase in the surface temperature of the star. 
This is consistent with the simultaneously increasing photometric magnitude. 
During nearly all observing seasons we see a high-latitude spot structure 
which is clearly nonaxisymmetric. The phase behaviour of this structure is 
very chaotic but agrees reasonably well with the photometry. 
Equatorial spots are also frequently seen, but many of them we interpret to 
be artefacts due to the poor to moderate phase coverage.  }
{Even during the chaotic phase of the star, the spot topology has remained very 
similar to the higher activity epochs with more coherent and long-lived spot structures: 
we see high-latitude and equatorial spot activity, 
the mid latitude range still being most often void of spots. 
We interpret the erratic jumps and drifts in phase of the photometric minima to be    
caused by changes in the high-latitude spot structure rather than the equatorial spots.}
 
\maketitle 

\section{Introduction}
\label{sec:intro}

LQ~Hya (HD 82558, GL 355) is a rapidly rotating single 
K2V star in the thin disk population \citep{fekel1986,Montes2001,Hinkel2017}.
LQ~Hya has an estimated mass of $0.8{\rm M}_\odot$ \citep{Kovari2004,tetzlaff2011},
an effective temperature of about 5000\,K \citep{donati1999,Kovari2004,Hinkel2017},
an estimated age from Lithium abundance of $51.9\pm 17.5$ Myr \citep{tetzlaff2011},  
and a measured rotation period of $\sim 1.60$ days 
\citep{fekel1986,jetsu1993,strassmeier1997,berdyugina2002,
Kovari2004,lehtinen2012,Olspert2015}.
Based on the spectral class and age, LQ~Hya is a young solar analogue and 
thus can provide insight into the dynamos of young solar-like stars.

Rapidly rotating convective stars are expected to generate magnetic fields 
through a dynamo process \citep[e.g.,][]{Berdyugina2005}.
The magnetic activity of LQ~Hya manifests as 
changes in the photometric light curve and chromospheric line emission. 
Variations in photometry of about 0.1 magnitudes 
and strong \ion{Ca}{II} H\&K emission lines indicative of chromospheric activity 
have been measured by \cite{fekel1986}, who classified LQ~Hya
as a BY-Draconis-type star as defined by \cite{Bopp1973}.
The changes in magnitude are thought to be due to 
starspots rotating across the line of sight with the stellar surface. 
These starspots are thought as analogues to sunspots, but generally 
large enough to decrease the stellar irradiance, unlike solar activity 
which is correlated with an increase in irradiance \citep[e.g.,][]{Yeo2014}.
These variations in stellar brightness and chromospheric emission 
are roughly cyclical, similar to the well-known 11-year cycle
seen in the sunspot number.

\begin{figure}
\includegraphics[width=\columnwidth]{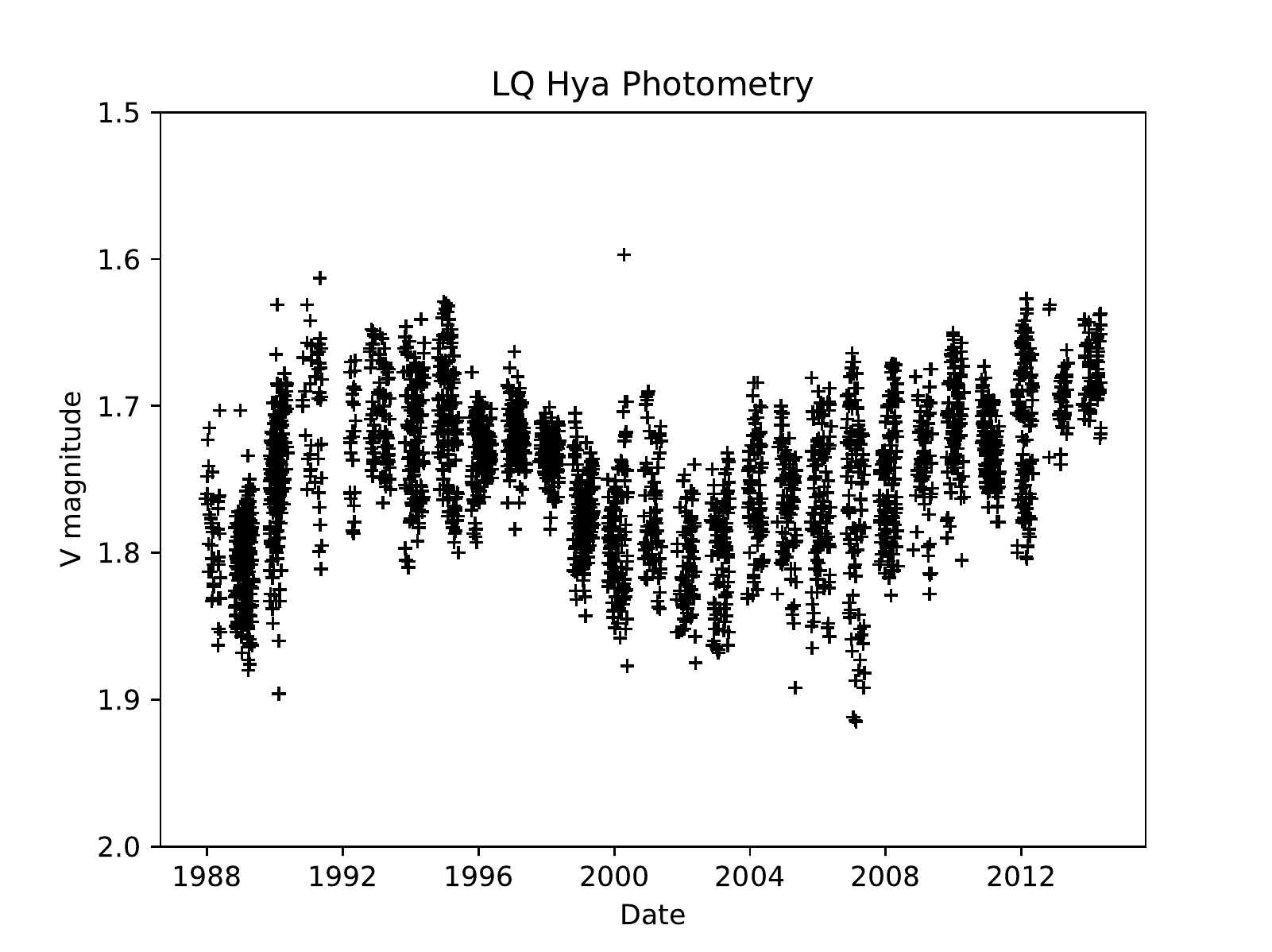}
\caption{V-band differential photometry of LQ~Hya from the T3 0.4\,m 
Automatic Photoelectric Telescope (APT) at the Fairborn Observatory, Arizona.
Data from \cite{Lehtinen2016}}\label{fig:photometry}
\end{figure}

Photometric studies spanning decades can be used to examine the periodicities 
in the light curve of LQ~Hya utilising time series analysis techniques.
\cite{jetsu1993} used a decade of photometry and found an overall
cycle period of 6.2 years for the mean brightness.
Changes in magnitude also correlated with changes in the observed
effective temperature based on the mean $B-V$ colour index. 
\cite{strassmeier1997} found a similar cycle fit of about 7 years.
With a longer timespan of photometric observations,
multiple cycles of 11.4, 6.8, and 2.8 years were observed by
\cite{olah2000} and cycles of 15 and 7.7 years were reported
by \cite{berdyugina2002}.
However, the longer the baseline of observations used, the less 
certain these cycles become as the activity appears to be somewhat chaotic.
Only weak indications of a 13 year cycle were detected by \cite{lehtinen2012} 
and a weak indication for a 6.9 year cycle was found by \cite{Olspert2015}.
\cite{Lehtinen2016} suggested a long and seemingly non-stationary cycle
with a period somewhere between 14.5 and 18 years as well as some indication
for 2 to 3 year oscillatons that could not be demonstrated to be periodic.
\cite{olah2009} revisited the photometric observations of their previous paper 
but with a longer dataset and found a 7 year cycle that steadily increased 
to 12.4 years when a longer timespan of photometry was used. 
From Figure \ref{fig:photometry}, we can see that cycle periods 
are difficult to estimate from the limited dataset because the longer cycle 
estimates are still a significant fraction of the total timespan of data.

Another phenomenon of interest that can be obtained from light 
curves is the existence of active longitudes, or the tendency of starspots
to occur at the same longitude for several years. 
These active longitudes may sometimes suddenly switch by 
about $180\degr$, commonly referred to as a flip-flop
\citep{Jetsu1993b}.
\cite{berdyugina2002} found two active longitudes approximately 
$180\degr$ apart for the duration of their photometric data with 
a phase drift of about $-0.012$ yr$^{-1}$ in the rotation frame.
\cite{lehtinen2012} detected only one stable active longitude between 
2003 and 2009 from their dataset while any other active longitudes 
were only stable for half a year. 
This is supported by the carrier fit analysis of 
\cite{Olspert2015} which finds the rotation periods of spots 
with linear trends between the epochs 1990--1994, and again for 2003--2009.
\cite{lehtinen2012} found possible flip-flops in late 1988, 1994, 1999,
2000, and 2010. \cite{Olspert2015} found agreement with the 1988, 1999,
and 2010 flip-flops and an additional possible flip-flop in late 1997.

Photometry mainly contains longitudinal and stellar magnitude information
and can only barely distinguish between high and low latitudes for good datasets 
\citep[e.g.,][]{berdyugina2002}. 
In order to study the spot latitudes, inversion methods must be applied 
to stellar spectra.
The Doppler imaging technique (hereafter DI) provides both latitudinal and 
longitudinal information for cool spots. 
Using this technique for LQ~Hya,
\cite{Strassmeier1993} found spotted regions mainly near the pole and 
equator with larger spot structures extending to mid-latitudes during 1991.
\cite{rice1998} found similar results from their maps, although the near-polar 
spots were weaker during 1993 and 1995. 
\cite{Kovari2004} recovered spots at low- to mid-latitudes for observations during 
1996 and 2000 and found the spot evolution to be rapid.
They speculated this to be the result of changes in the emergence rate of 
magnetic flux and not spot migration.
\cite{cole2014a} found some evidence for a bimodal structure with spots either at 
high or low latitudes for seven observing seasons spanning from 1998--2002. 
\cite{Soriano2017} reconstructed temperature maps from late 2011 to mid-2012 and 
found two large near-polar spots and one low-latitude spot that migrated 
equatorward over the course of the observations. 

The Zeeman Doppler imaging technique (hereafter ZDI) yields similar results of bimodal 
structure with spots either at very low or very high latitudes.
Spot occupancy maps from 1991--2002 from \cite{donati1999} and \cite{donati2003} 
reveal near-polar spots for some observing seasons and spots between the equator and 
$\pm30\degr$ latitudes corresponding to the radial and azimuthal magnetic field components 
with components measuring as high as 900\,G 
and a mean quadratic magnetic field flux between 30--100\,G. 
Furthermore, they found that the assumed relationship of \cite{berdyugina2002}
between dark, low latitude spots and photometric minima to not be upheld by the ZDI results, 
but rather, dependent on multiple phenomena such as the nonaxissymetric 
polar features.
\cite{donati2003} also observed that the spot evolution is not the result of equatorward 
drift of spots from high to low latitudes, but rather seems to the result of 
some other mechanism that causes high and low latitude spots to form at 
varying strengths over time. 

The surface differential rotation of LQ~Hya appears to be small.
Estimates from photometry range from $k=0.015$--$0.025$ where 
$k = (\Omega_{\textrm{eq}}-\Omega_{\textrm{pol}})/\Omega_{\textrm{eq}}$
and $\Omega_{\textrm{eq}}$ is the rotation rate at the equator and 
$\Omega_{\textrm{pol}}$ is the rotation rate at the poles
\citep{jetsu1993,You2007,Olspert2015}.
\cite{berdyugina2002} tracked active longitudes and 
estimated a differential rotation rate of $k=0.002$ based on the period differences.
Estimates from DI give a similarly small 
differential rotation rate of $k=0.006$ \citep{Kovari2004},
and \cite{donati2003b} find from ZDI maps that the measured differential rotation 
switches 
between almost solid body rotation ($k=0.003$) and weak differential rotation ($k=0.05$) .
Thus, the differential rotation measurements are not conclusive, but all reported 
results point to a small $k$-value.
Hence, in this study we do not include differential rotation into the inversion procedure.

Evidently, the star seems highly variable in its spot activity, 
with periods of long-lived spot structure 
and periods of chaotic and rapid spot evolution.
As \cite{Lehtinen2016} note, from the photometry alone it becomes apparent
that while LQ~Hya was cyclical for earlier epochs, it seems now to be steadily 
increasing in surface brightness with no overt signs of 
stopping (see Figure~\ref{fig:photometry}).
The only exception to this upward trend is 
a slight dip in the curve  
downward between 2009--2011.
Such a trend would indicate 
an increase in the magnetic activity level of the star.
This paper aims to examine the spot topology of LQ~Hya from 2006--2017 using the 
DI technique during this period of decreasing activity level.

\section{Data} \label{sec:DATA}

\begin{table*}
\tiny
\caption{List of all observations}
\centering
\begin{tabular}{ccccc|ccccc}
\hline \hline
Instrument & Date & HJD &$\phi$ & S/N & Instrument & Date & HJD &$\phi$ & S/N \\
 &(dd/m/yyyy)&-2 400 000&       &     &  &(dd/m/yyyy)&-2 400 000&       &     \\       
\hline
SOFIN  &  02/12/2006  &  54071.7383  &  0.548  &  168  &   SOFIN  &  13/12/2011  &  55908.7461  &  0.863  &  418 \\  
SOFIN  &  03/12/2006  &  54072.7578  &  0.185  &  289  &   SOFIN  &  14/12/2011  &  55909.6992  &  0.458  &  434 \\
SOFIN  &  04/12/2006  &  54073.7617  &  0.812  &  185  &   SOFIN  &  15/12/2011  &  55910.7305  &  0.102  &  224 \\
SOFIN  &  05/12/2006  &  54074.7422  &  0.424  &  221  &   SOFIN  &  23/11/2012  &  56254.7461  &  0.960  &  360 \\
SOFIN  &  06/12/2006  &  54075.7461  &  0.051  &  214  &   SOFIN  &  28/11/2012  &  56259.7539  &  0.087  &  298 \\
SOFIN  &  23/11/2007  &  54427.7695  &  0.909  &  215  &   SOFIN  &  04/12/2012  &  56265.7500  &  0.832  &  333 \\
SOFIN  &  27/11/2007  &  54431.7695  &  0.408  &  179  &   SOFIN  &  05/12/2012  &  56266.7539  &  0.459  &  369 \\
SOFIN  &  28/11/2007  &  54432.7734  &  0.035  &  272  &   SOFIN  &  15/11/2013  &  56611.7461  &  0.926  &  229 \\
SOFIN  &  01/12/2007  &  54435.7695  &  0.906  &  217  &   SOFIN  &  20/11/2013  &  56616.7695  &  0.064  &  196 \\
SOFIN  &  02/12/2007  &  54436.7695  &  0.530  &  268  &   SOFIN  &  21/11/2013  &  56617.7656  &  0.686  &  336 \\
SOFIN  &  03/12/2007  &  54437.7773  &  0.160  &  217  &   SOFIN  &  22/11/2013  &  56618.7656  &  0.310  &  362 \\ \cline{6-10}
SOFIN  &  09/12/2008  &  54809.7031  &  0.449  &  422  &   FIES   &  03/12/2014  &  56994.7070  &  0.107  &  280 \\
SOFIN  &  10/12/2008  &  54810.7383  &  0.095  &  205  &   FIES   &  05/12/2014  &  56996.6563  &  0.325  &  290 \\
SOFIN  &  11/12/2008  &  54811.6992  &  0.695  &  297  &   FIES   &  07/12/2014  &  56998.7070  &  0.605  &  215 \\
SOFIN  &  12/12/2008  &  54812.7148  &  0.330  &  251  &   FIES   &  08/12/2014  &  56999.7148  &  0.235  &  210 \\
SOFIN  &  15/12/2008  &  54815.7578  &  0.230  &  67   &   FIES   &  26/11/2015  &  57352.7656  &  0.735  &  60  \\ 
SOFIN  &  27/12/2009  &  55192.6953  &  0.649  &  316  &   FIES   &  27/11/2015  &  57353.7227  &  0.333  &  180 \\
SOFIN  &  30/12/2009  &  55195.6914  &  0.520  &  317  &   FIES   &  28/11/2015  &  57354.7227  &  0.957  &  190 \\
SOFIN  &  31/12/2009  &  55196.6758  &  0.135  &  223  &   FIES   &  30/11/2015  &  57356.7227  &  0.206  &  180 \\
SOFIN  &  01/01/2010  &  55197.6836  &  0.764  &  373  &   FIES   &  03/12/2015  &  57359.6914  &  0.061  &  240 \\
SOFIN  &  05/01/2010  &  55201.6328  &  0.231  &  304  &   FIES   &  19/12/2017  &  58106.6914  &  0.604  &  170 \\
SOFIN  &  14/12/2010  &  55544.6797  &  0.483  &  330  &   FIES   &  20/12/2017  &  58107.5547  &  0.143  &  210 \\
SOFIN  &  23/12/2010  &  55553.7461  &  0.146  &  376  &   FIES   &  20/12/2017  &  58107.7734  &  0.280  &  240 \\
SOFIN  &  24/12/2010  &  55554.7305  &  0.760  &  405  &   FIES   &  21/12/2017  &  58108.6953  &  0.856  &  220 \\
SOFIN  &  25/12/2010  &  55555.7461  &  0.395  &  351  &   FIES   &  22/12/2017  &  58109.5859  &  0.412  &  120 \\
SOFIN  &  26/12/2010  &  55556.7656  &  0.031  &  283  &   FIES   &  22/12/2017  &  58109.7148  &  0.493  &  180 \\
SOFIN  &  09/12/2011  &  55904.7305  &  0.355  &  363  &   FIES   &  23/12/2017  &  58110.5977  &  0.044  &  240 \\
SOFIN  &  11/12/2011  &  55906.7188  &  0.597  &  251  &   FIES   &  23/12/2017  &  58110.7383  &  0.132  &  290 \\ 
SOFIN  &  12/12/2011  &  55907.7422  &  0.236  &  293  \\   

\hline
\end{tabular}
\tablefoot{HJD is -2 400 000.}\label{table:fullobs}
\end{table*}

We have collected 
11 sets of winter-season spectra, covering the time interval 2006--17, with
the 2.56\,m Nordic Optical Telescope at La Palma, Spain.
The details of the observations are given in Table~\ref{table:fullobs} and the season summaries  
in Table~\ref{table:obs}. 
For the years 2006--2013 we used the SOFIN instrument, which is a high-resolution
\'echelle spectrograph mounted in the Cassegrain focus,
while for the years 2014--2017 we used FIES, which is a fiber-fed \'echelle
spectrograph. The former has also a spectropolarimetric mode available, but in this paper we
only analyse the unpolarised spectroscopy obtained with the two instruments, the aim being to monitor
the behaviour of starspots in terms of temperature anomalies on the stellar surface.
The SOFIN observations were reduced with the new SDS tool,
which is described in some detail in \cite{Willamo2019}. 
The FIES observations were reduced with the standard
FIEStool pipeline \citep{Telting2014}.
The spectral resolutions of SOFIN and FIES data sets are 70 000 and 67 000, respectively.
The observations have mostly  
poor to moderate phase coverage
of 39--68\%, while the signal-to-noise 
ratio (hereafter S/N) is reasonably good
with mean values exceeding 200 for all but one season. 

For phasing the observations, we used the rotation period and the ephemeris from \cite{jetsu1993},
\begin{equation}
{\rm HJD}_0 = 2 445 274.22  + 1.601136 {\rm E}, \label{ephemeris}
\end{equation}
where $\rm HJD_0$ corresponds to the zero phase. Other stellar parameters adopted, 
listed in Table~\ref{table:stellarparameters}, closely follow those of
\cite{cole2014b}, except for the surface gravity and microturbulence values. 
For the former, we used a more standardly reported value of $\log{g}=4.5$ 
\citep{Tsantaki2014},
while for latter, we adopted a somewhat higher value of $\xi_{\rm t}$ = 1.5 km s$^{-1}$, 
found by 
fitting the mean spectral lines
from all seasons 
of our data to a model calculated for an unspotted surface.

The spectral regions 6438.4--6439.8\,\AA, 6461.8--6463.5\,\AA,
and 6471.0--6472.4\,\AA\, were used for SOFIN observations, while for FIES, three
additional spectral regions of 
6410.9--6412.4\,\AA, 6419.3--6422.0\,\AA, and 6430.2--6431.5\,\AA\, were used. 
The FIES spectral regions overlap both with those used by \cite{cole2014b} and the 
SOFIN observations in this paper.

\begin{table}
\tiny
\tiny
\caption{Summary of the observing seasons}
\centering
\begin{tabular}{cccccc}
\hline \hline
Time &Instrument &$n_{\phi}$ &$\langle$S/N$\rangle$ &$f_{\phi}$ [\%] &$\sigma$[\%]\\
\hline
Dec 2006 &SOFIN &5 &215 &50 &0.542\\  
Dec 2007 &SOFIN &6 &228 &53 &0.491\\ 
Dec 2008 &SOFIN &5 &248 &50 &0.936\\ 
Dec 2009 &SOFIN &5 &306 &50 &0.388\\ 
Dec 2010 &SOFIN &5 &344 &49 &0.412\\ 
Dec 2011 &SOFIN &6 &268 &60 & 0.448 \\  
Dec 2012 &SOFIN &4 &340 &40 & 0.380\\  
Nov 2013 &SOFIN &4 &281 &40 &0.496\\ \hline  
Dec 2014 &FIES  &4 &249 &39 &0.425\\ 
Dec 2015 &FIES  &5 &170 &50 &0.637\\ 
Dec 2017 &FIES  &8 &209 &68 &0.549\\ \hline 
\end{tabular}
\tablefoot{The number of phases in each observing season is given by $n_{\phi}$,
the mean signal to noise ratio by $\langle$S/N$\rangle$, and the phase coverage by $f_{\phi}$, which
was computed assuming a phase range of $\phi \pm 0.05$ for each observation. 
We also list the deviation, $\sigma$,
of the inversion solution compared to the observations.}\label{table:obs}
\end{table}

\section{Doppler imaging} \label{sec:DI}

\begin{table}
\tiny
\caption{Adopted stellar parameters}
\label{table:stellarparameters}
\centering
\begin{tabular}{rl}
\hline \hline
Parameter &Value\\
\hline
Effective temperature &$T_{\rm eff} $ = 5000\,K\\
(unspotted) &\\
Gravity                        &log $g$ = 4.5                 \\
Inclination                   &$i$ = 65$^{\circ}$           \\
Rotational velocity      &$v\,{\rm sin}\, i$ = 26.5 km s$^{-1}$  \\
Rotation period           &$P$ = 1${\fd}$601136        \\
Metallicity                    &log[M/H] = 0                    \\
Microturbulence         &$\xi_{\rm t}$ = 1.5 km s$^{-1}$ \\
Macroturbulence          &$\zeta_t$ = 1.5 km s$^{-1}$  \\ \hline
\end{tabular}
\tablefoot{All other values of stellar parameters chosen from \cite{cole2014b} except
for surface gravity and microturbulence values.
Surface gravity is the same as in \cite{Tsantaki2014} and the 
microturbulence is fitted to the data. 
}\label{star}
\end{table}

To invert the observed spectroscopic line profiles into a surface temperature 
distribution on the stellar surface, we used the DI code INVERS7DR \citep[see e.g.][]{Willamo2019}, 
that uses Tikhonov regularisation for the ill-posed inversion problem 
\citep{Piskunov1990}. 
The regularisation technique minimises temperature gradients in the solution 
and hence tends to damp small-scale features.

To construct a model spectrum for the star, we retrieved the spectral parameters from the
Vienna Atomic Line Database \citep{Kupka1999,Ryabchikova2015}, 
using effective temperatures of 
5000\,K and 4000\,K for the unspotted and spotted stellar surface, respectively.
We used a total of 114 lines for the SOFIN spectral regions, and 221 lines for the FIES spectral regions
for the computation of the synthetic spectra. 
Line profiles were calculated using plane-parallel $\log g = 4.5$ stellar atmosphere
models taken from the MARCS database \citep{Gustafsson2008}.
The lines used for inversions are \ion{Fe}{I} and \ion{Ca}{I} lines. 
We assume solar metallicity and adjust individual spectral lines to fit the mean observations. 
The stellar lines used and their original and adopted parameters 
are listed in Table~\ref{table:lineparameters}. 
The \ion{Ca}{I} lines are susceptible to 
non-local thermal equilibrium (NLTE) effects in the temperature range of LQ~Hya,
but a simple test excluding these lines from the inversion procedure 
did not alter the results significantly.
The NLTE effects were likely mitigated by our use of a higher value for 
${\rm log}(gf)$.
The models covered the temperature range 3500--6000 K.

\begin{table}
\tiny
\caption{Parameters for the absorption lines used in the inversion.}
\label{table:lineparameters}
\centering
\begin{tabular}{ccc}
\hline \hline
Line [\AA] &${\rm log}(g f )$ &${\rm log}(g f )_{\rm standard}$\\
\hline
\ion{Fe}{I} $6411.6476$ &$-0.675$ & $-0.595$ \\
\ion{Fe}{I} $6419.9483$ &$-0.300$ & $-0.240$ \\
\ion{Fe}{I} $6421.3495$ &$-2.250$ & $-2.027$ \\
\ion{Fe}{I} $6430.8446$ &$-2.050$ & $-2.106$ \\  
\ion{Ca}{I} $6439.0750$ &$0.400$ & $0.390$ \\
\ion{Fe}{I} $6462.7251$ &$-2.100$ & $-2.367$ \\ 
\ion{Ca}{I} $6471.6620$ &$-0.350$ &$-0.686$\\ \hline
\end{tabular}
\tablefoot{Adopted vs. standard VALD ${\rm log}(g f )$ values for the chosen lines. }
\end{table}

The surface grid resolution used for the inversion
was $40\times80$ in latitude and longitude, respectively. The inversion
was run with the regularisation parameter $2.5 \times 10^{-9}$ for 100 iterations, 
at which point a sufficient convergence was reached. 
The final deviation between the inversion solution and the observations
$\sigma[\%]$
is indicated in the sixth column in Table~\ref{table:obs}.
We constrained the inversion process by imposing lower and upper 
temperature limits of 3500\,K and 5500\,K, respectively. 
This was necessary because our observations generally 
have only a modest phase coverage and the inversions tend to produce 
features with very high temperatures as a result. 
Such features are not likely to be physically real for a 
cool star such as LQ~Hya and so we adopted the upper temperature limit. 
The temperature constraint was done by adding a penalty function to the 
minimisation procedure as described in
\cite{Hackman2001}.
In all cases this procedure was not 
observed to change the overall topology of the solution.

\section{Results} \label{sec:RESULT} 

\begin{figure*}
\includegraphics[width=0.9\columnwidth]{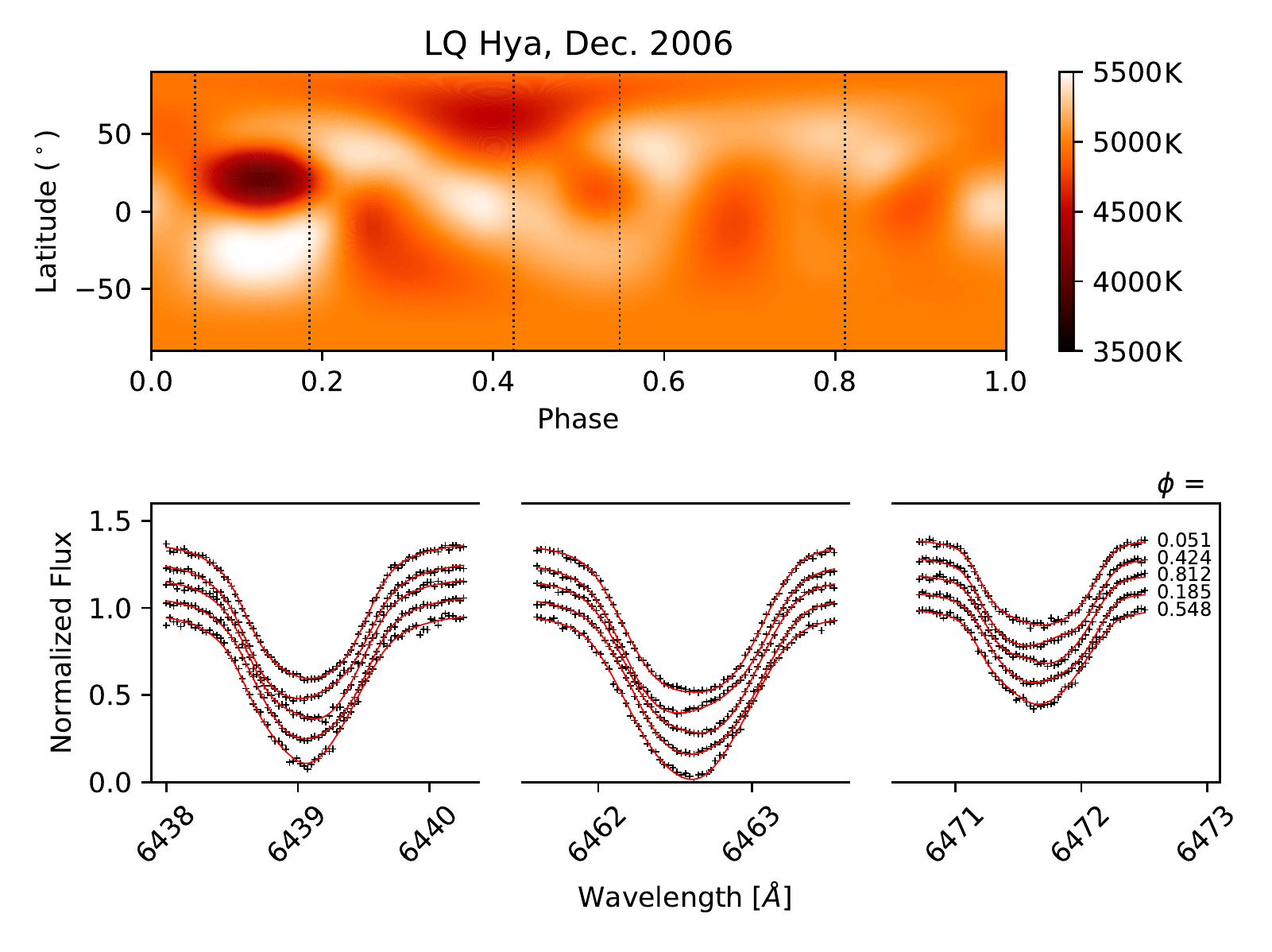}
\includegraphics[width=0.9\columnwidth]{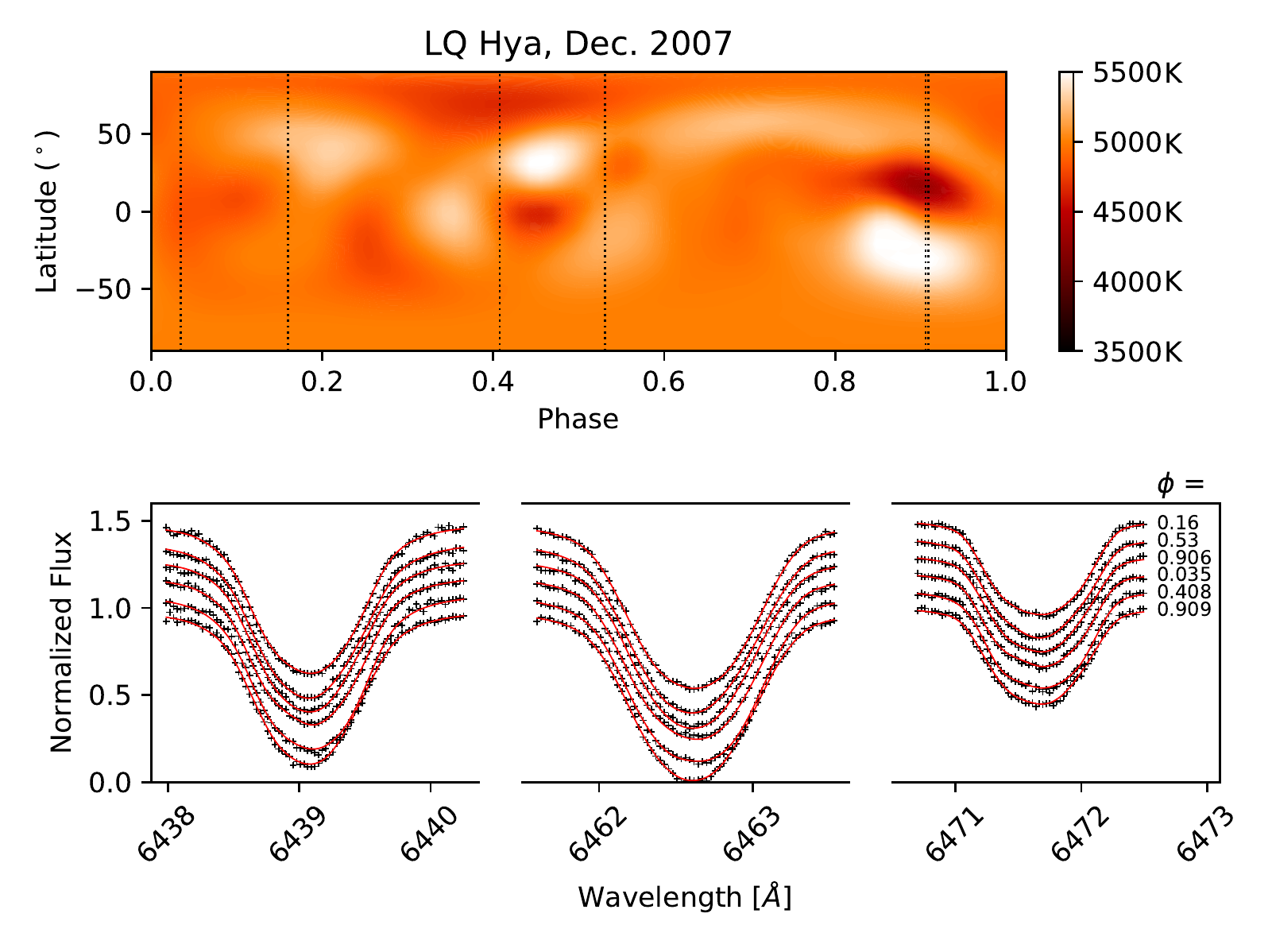}\\
\includegraphics[width=0.9\columnwidth]{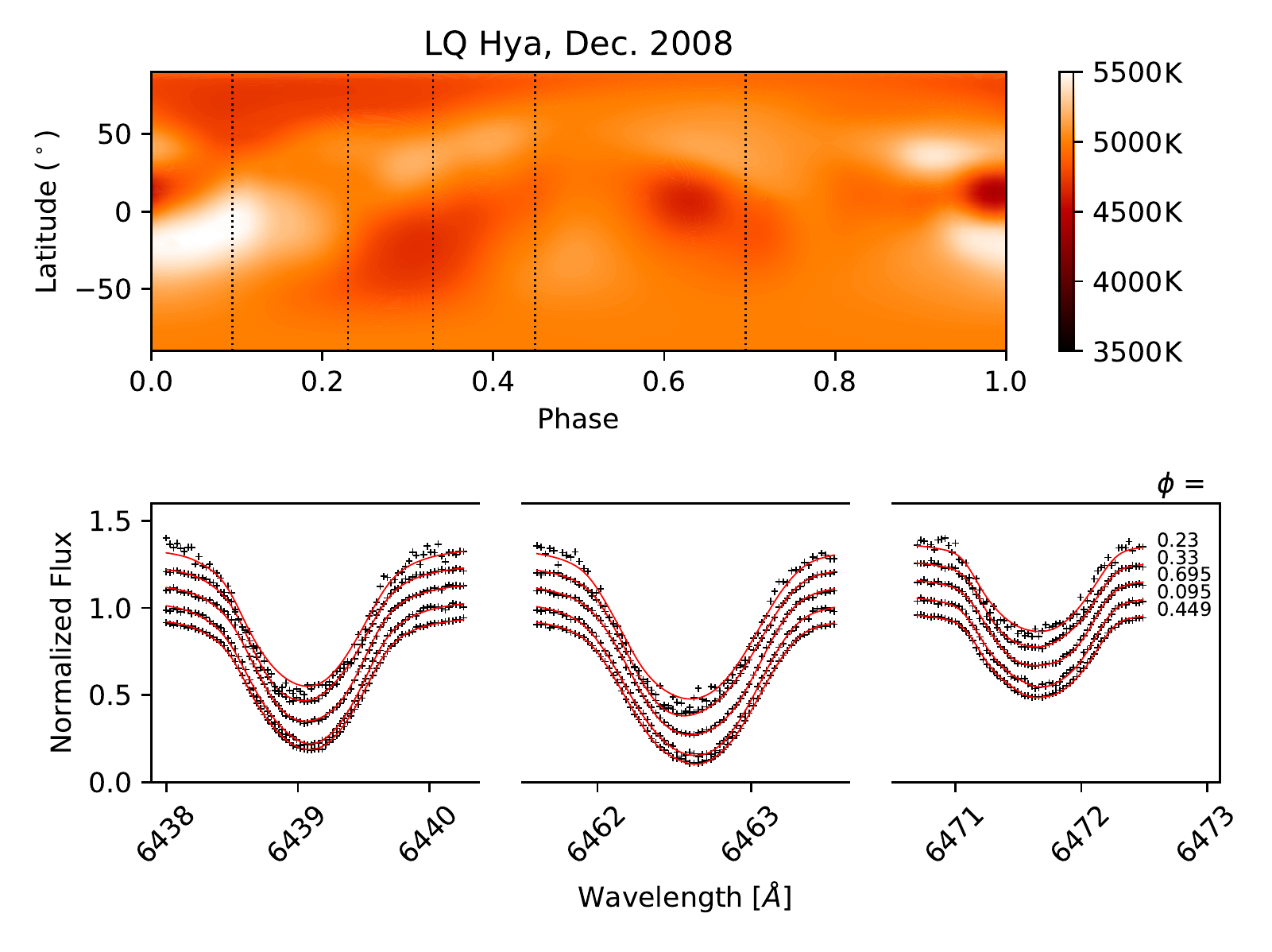}
\includegraphics[width=0.9\columnwidth]{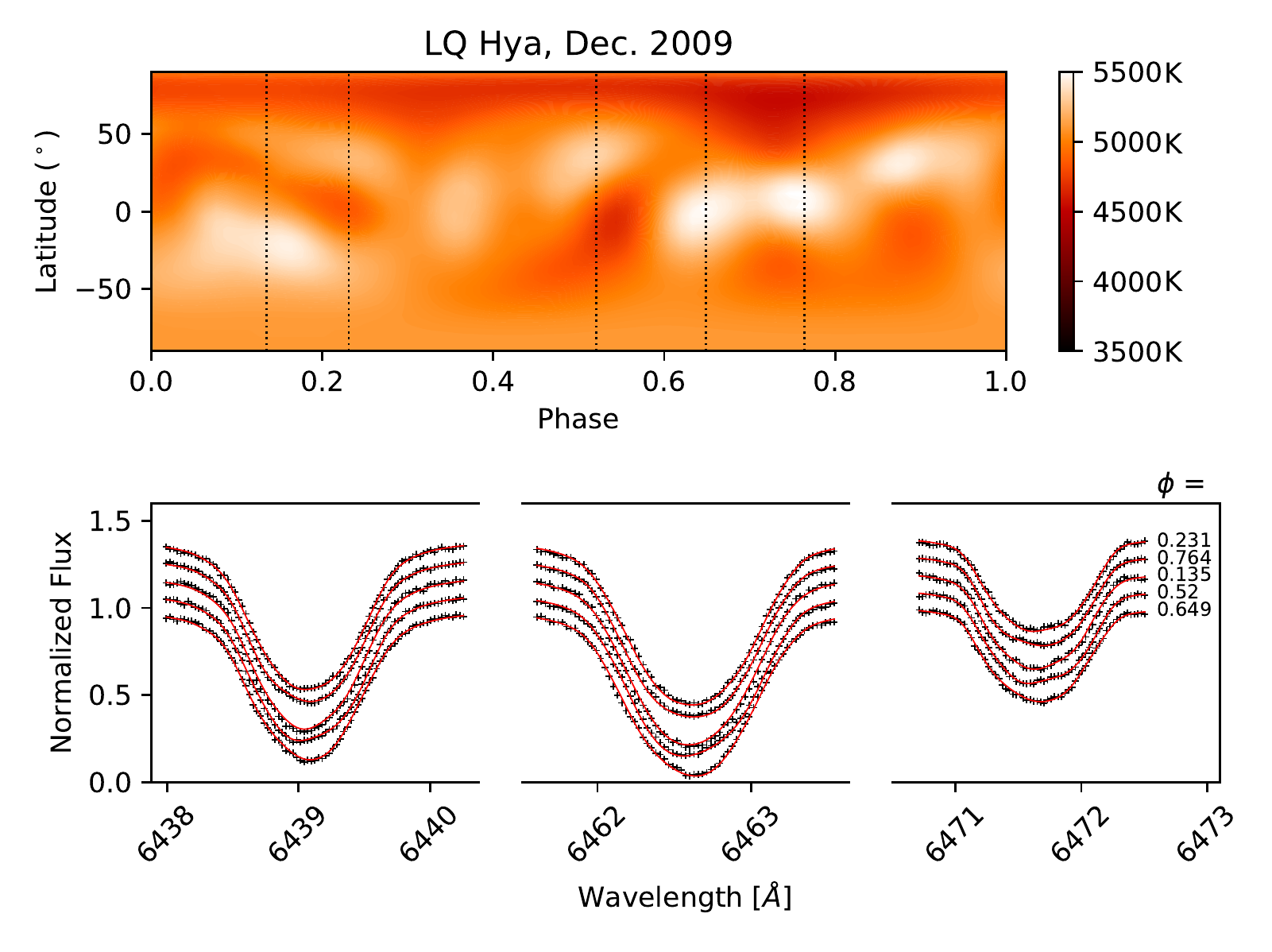}\\
\includegraphics[width=0.9\columnwidth]{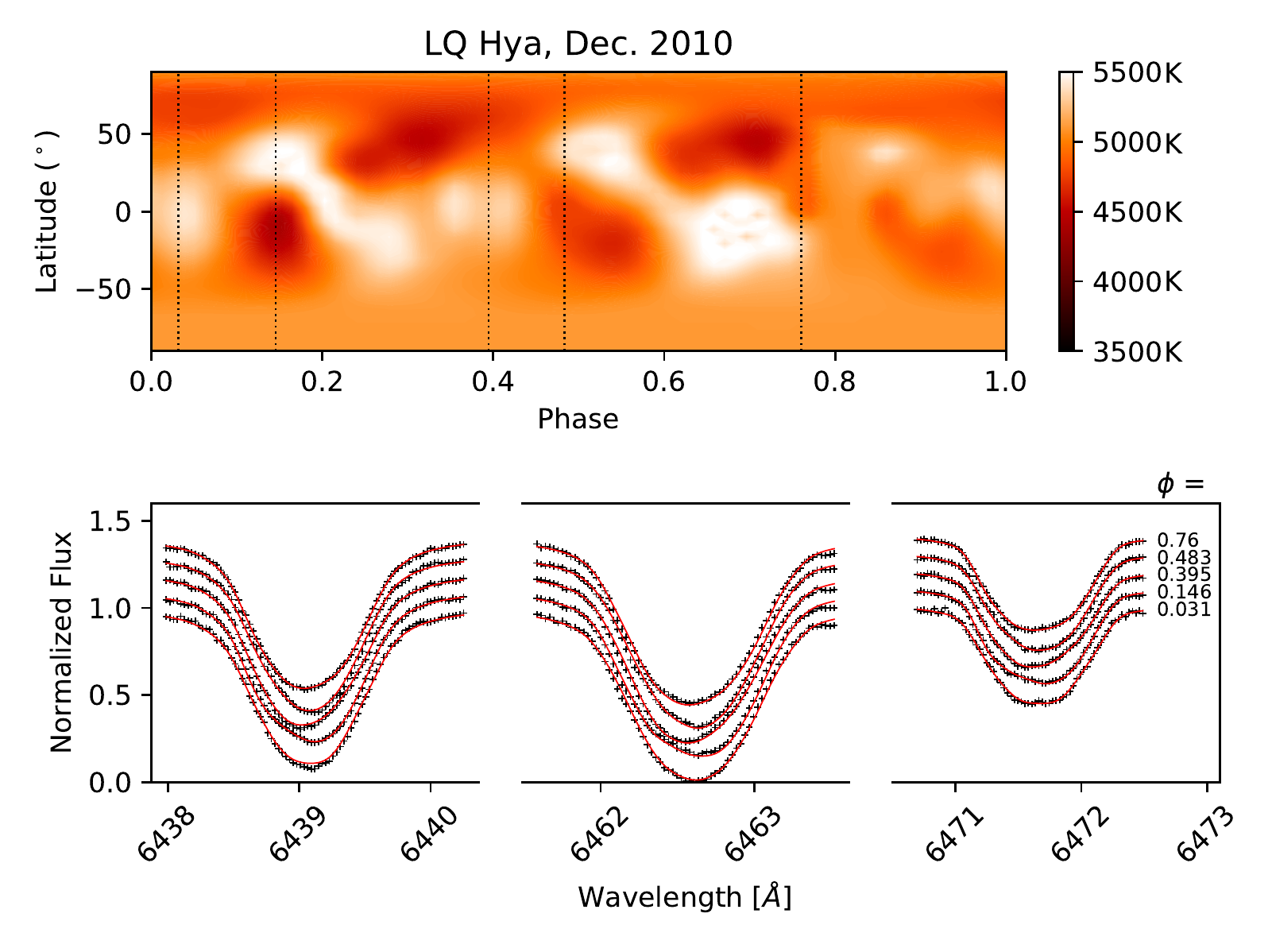}
\includegraphics[width=0.9\columnwidth]{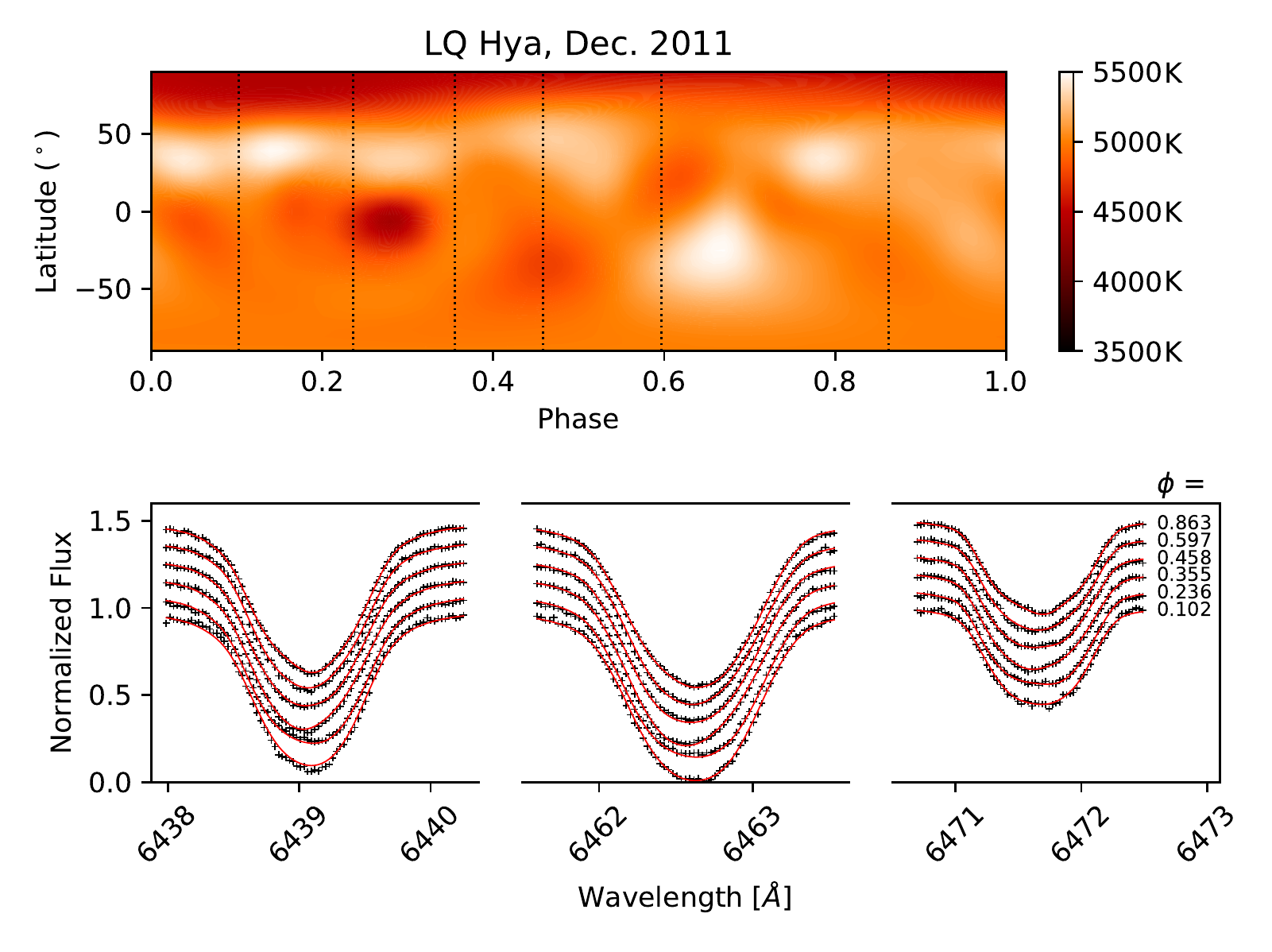}\\
\includegraphics[width=0.9\columnwidth]{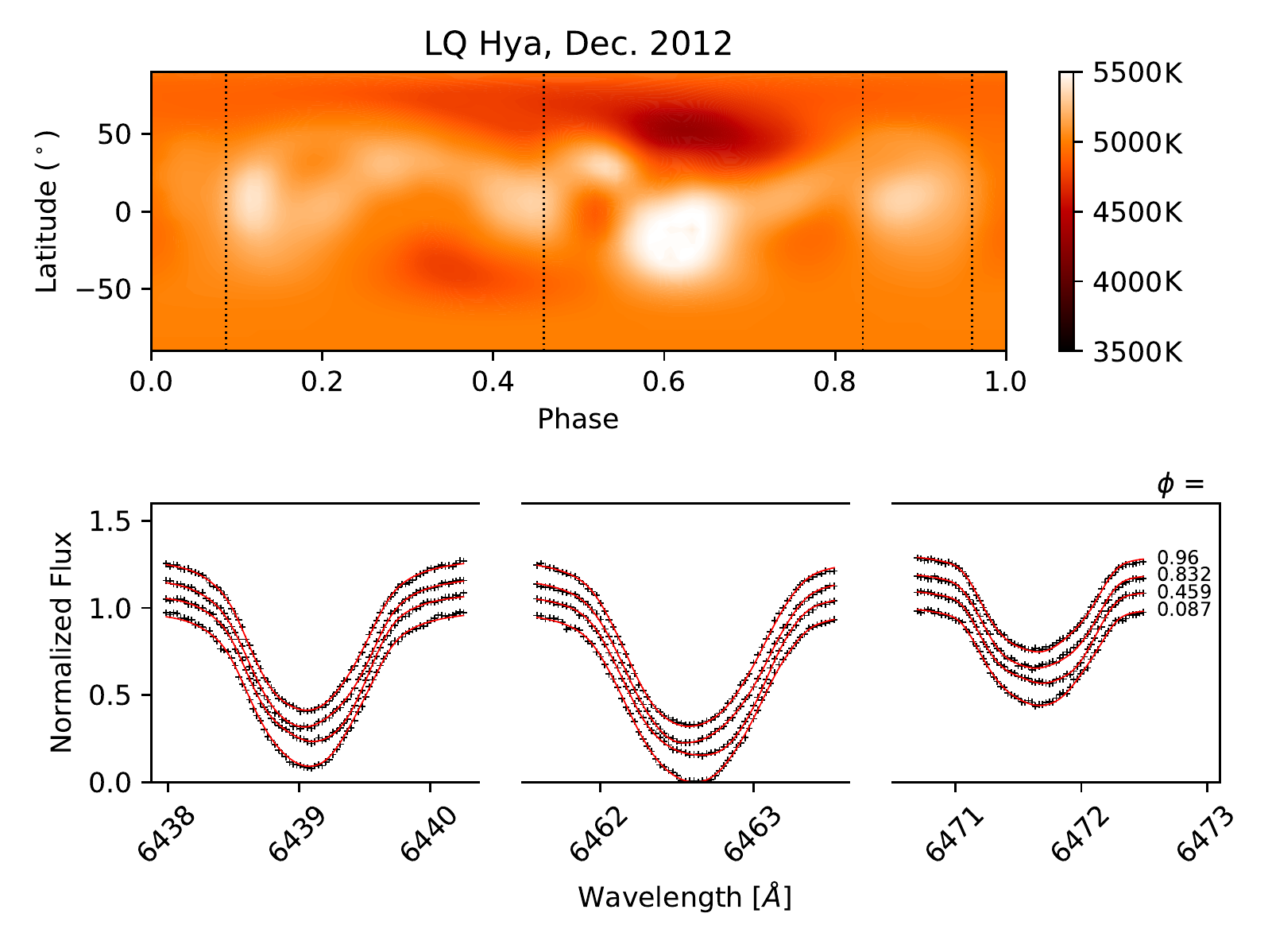}
\includegraphics[width=0.9\columnwidth]{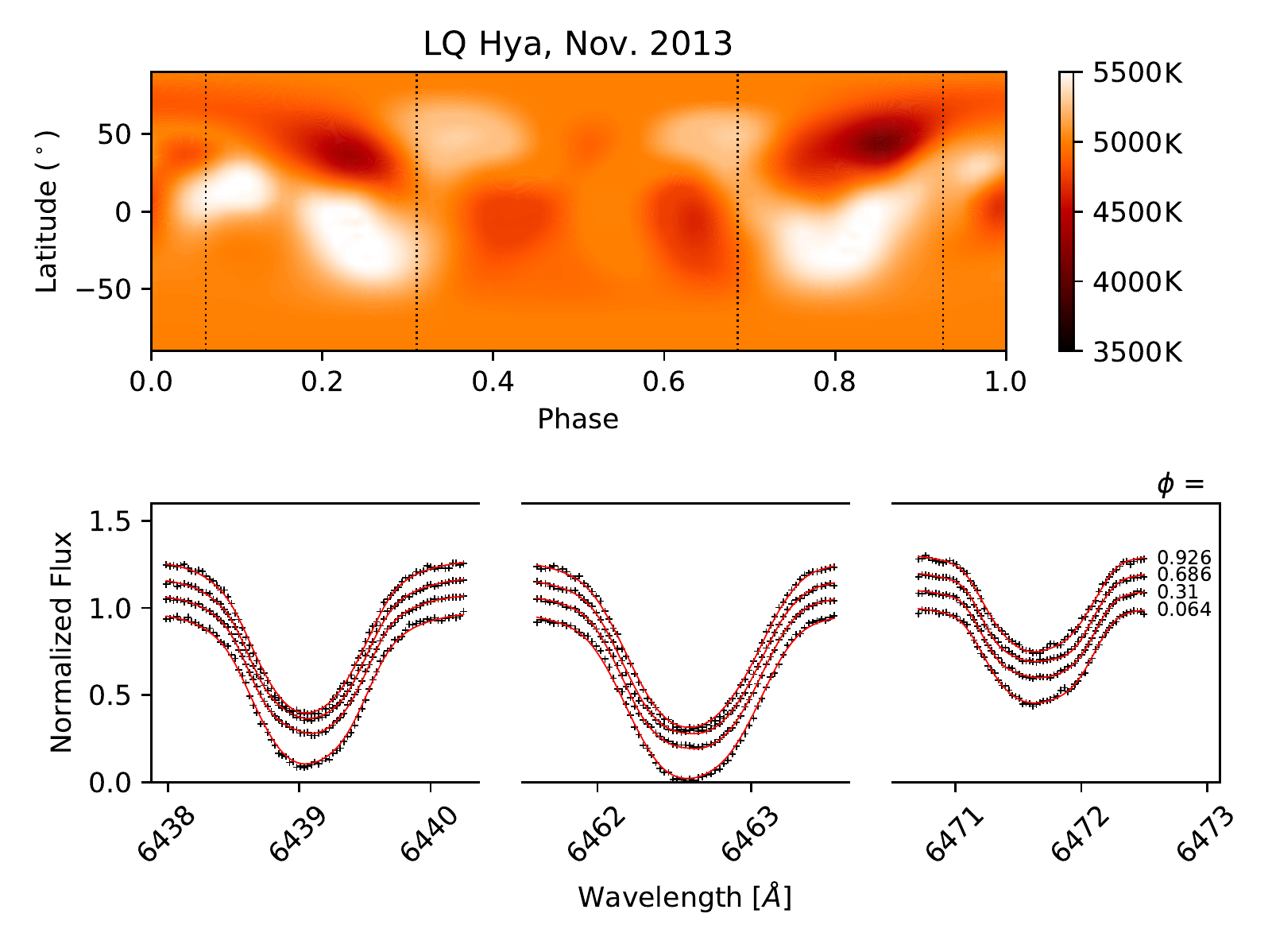}\\
\caption{Doppler images from SOFIN observations 2006--2013.}
\label{fig:di_set1}
\end{figure*}

\begin{figure*}
\includegraphics[width=0.9\columnwidth]{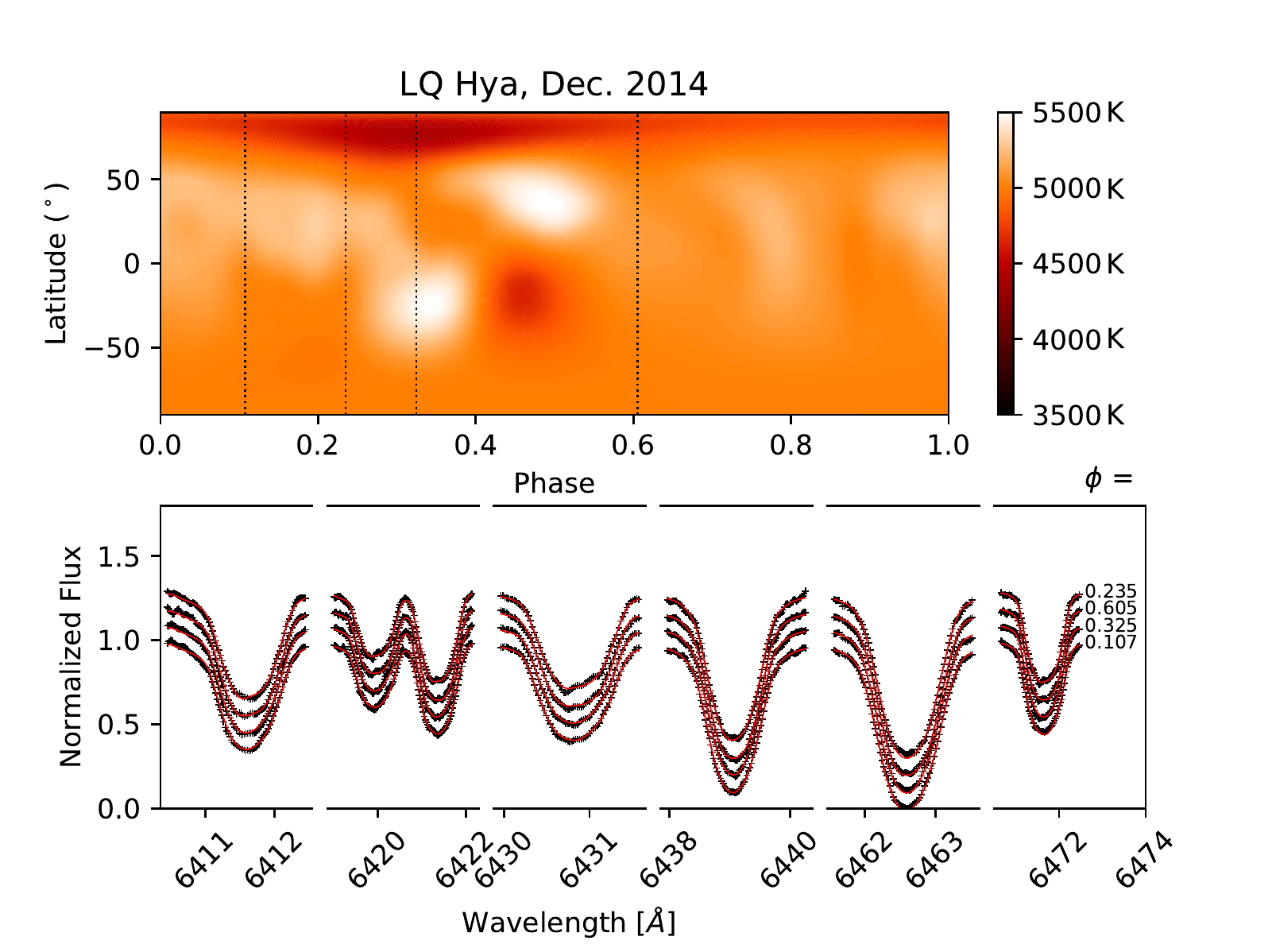}
\includegraphics[width=0.9\columnwidth]{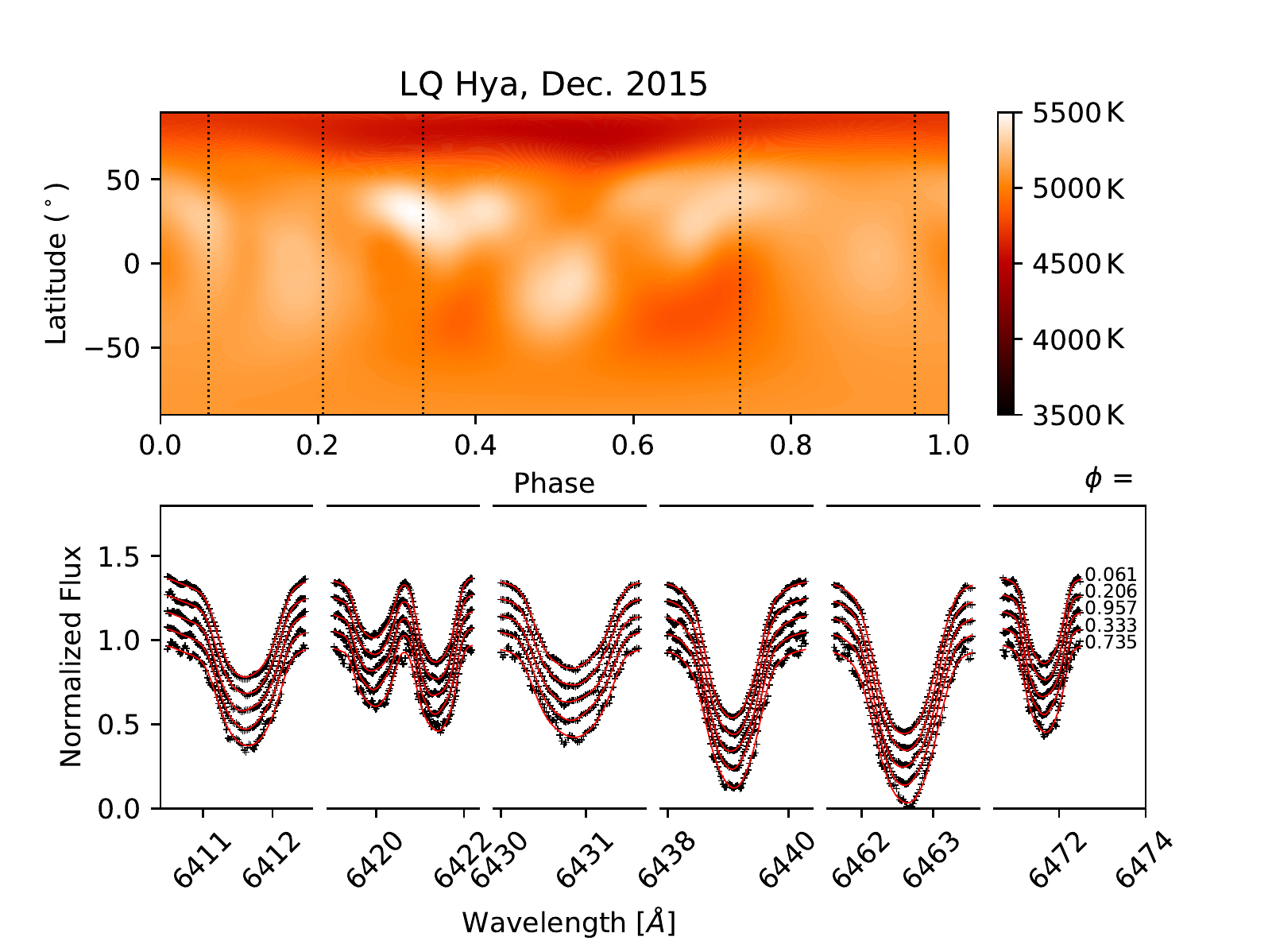}
\begin{center}
\includegraphics[width=0.9\columnwidth]{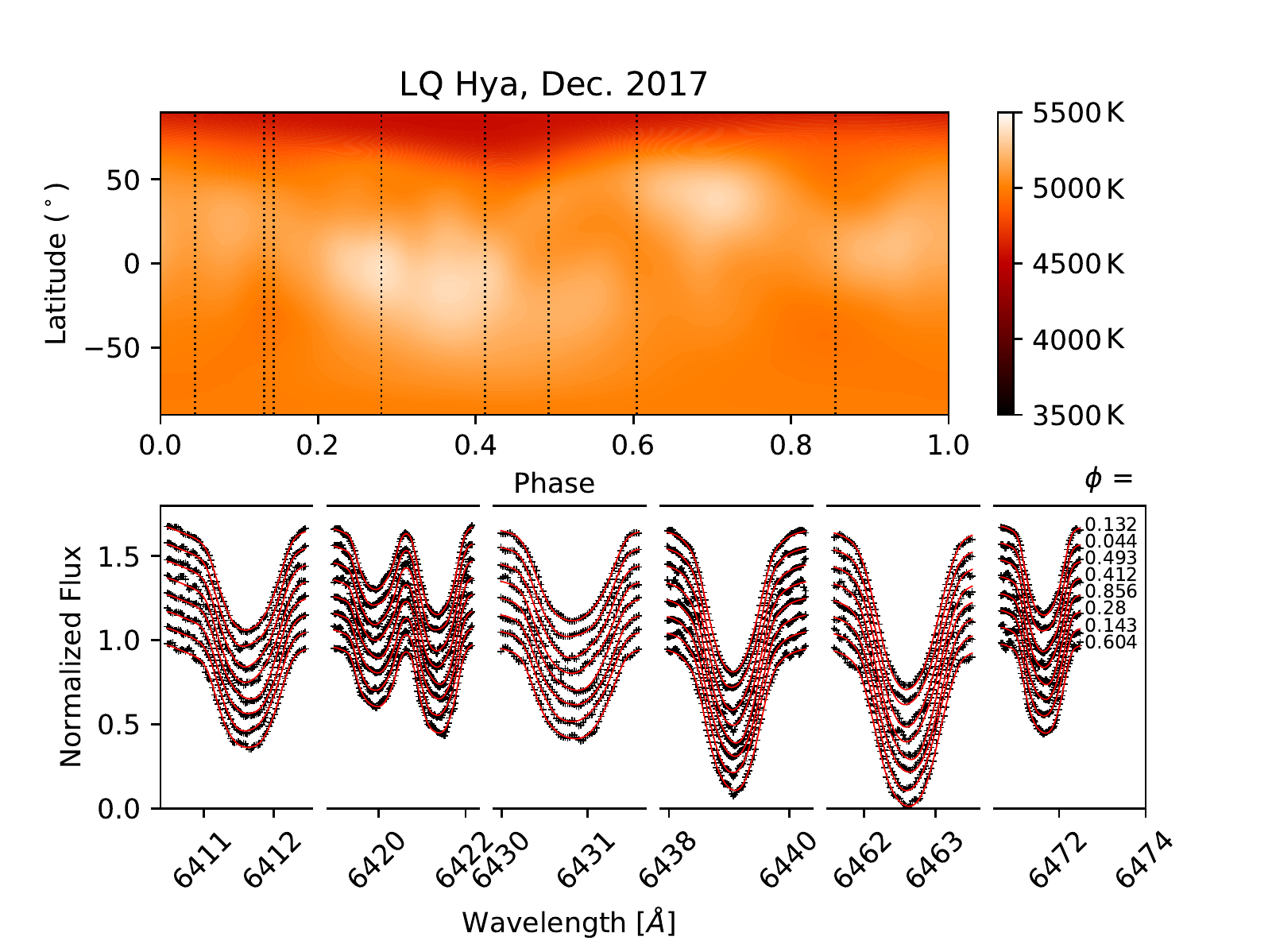}
\end{center}
\caption{Doppler images from FIES observations 2014, 2015, and 2017.}
\label{fig:di_set2}
\end{figure*}

\subsection{Temperature maps from SOFIN observations, 2006--2013}\label{SOFIN}

We use the DI technique and solve for the surface temperature for 
each of the observing season.
Our S/N is good for all but two of the individual observations, 
but all observations within a season are weighted based on 
their S/N so that noisier observations have less of an impact on the final map.  
Despite the good S/N, our results still need to be treated with some reservation 
because we have poor to moderate phase coverage for most of our seasons.
From Table \ref{table:obs} we can see that the best seasons are Dec.2011 and Dec.2017 
with a $f_\phi$ of 60\% and 68\% respectively,
where $f_\phi$ was calculated by assuming a phase range of $\phi \pm 0.05$ for each observation.
All other seasons have a $f_\phi$ of 53\% or less and thus interpretation 
of the maps should be treated carefully.
This will be addressed further in Section~\ref{section:discussion}.

Temperature maps for the SOFIN observations are presented in 
Figure~\ref{fig:di_set1}. 
\begin{itemize}
\item The Dec.2006 observing season has a 50\% phase coverage.
The cool spot near phase 0.12 is latitudinally paired with a hot spot, 
which are likely artefacts. 
There is evidence of a cool spot at high latitudes at phase around 0.4, 
but no evidence of a secondary high-latitude spot structure
at phase 0.8, which agrees with the photometric results of 
\cite{Olspert2015}: during 2006--2010, the 
spot evolution was primarily dominated by one spot 
that showed rather chaotic phase behaviour.  
\item The Dec.2007 observations have a phase coverage of 53\% and exhibit some artefacts 
in the area of the large phase gap of $135\degr$. 
We again find evidence for a large high-latitude spot around phase 0.4, hence the primary
spot structure seems to still appear at the same longitude as during the previous year, which is 
not in good agreement with photometry of  \cite{Olspert2015}, which indicates primary spot structure at around
phase 0.9 and a secondary spot at phase 0.4. 
There are also two low-latitude 
cool spots paired with hot 
spots at the same phase, but these might be artefacts. 
\item The Dec.2008 observations show cool spots both near the equator and near the poles but the phase 
coverage is 50\% and the observation at $\phi=0.2$ is noisy. 
This may cause the inversion program to produce artificial 
spots near that particular phase.
The primary high-latitude spot structure now appears close to phase 0, which agrees
better with the photometry of \cite{Olspert2015}. Of the equatorial spots the ones
close to phase 0.4 and 0.6 seem realistic, although there are weak hot shadows paired
with them at higher latitudes. 
\item 
In Dec.2009, with a phase coverage of 50\%, we get a very strong high-latitude
spot structure with a temperature minimum around the phase 0.75. 
The structure is elongated in phase, 
almost forming an asymmetric cool polar cap. 
The location of the temperature minimum matches well with photometry of \cite{Olspert2015}. 
Again, lower latitude features are abundant, but 
paired hot shadows at the same phase accompany most of them. In
between phases 0.6 and 1.0 we see a 4-leaf clover structure of two of these features, cool-hot
and hot-cool pair adjacent to each other. 
Such a feature can be caused by a spot close to the visible southern limb, but can also
be an artefact. 
\item The observations for Dec.2010 have a phase coverage of 49\% and 
the map seems to be dominated by artefacts with no evidence of high-latitude activity. 
We see a checkerboard pattern around the equator at all phases, most likely resulting
from the combination of poor phase coverage with a long observation period during which the star may have changed.
Just by inspecting the line profiles, one sees strong spot variability, but it is impossible
to judge which of the features in the map itself are real and which are artefacts. Photometry
indicates again one primary spot structure around phase 0.6, where no spectroscopy
is available. After this season, the star appears
to enter a very chaotic state, characterised by frequent phase jumps, which were classified as 
flip-flops
by \cite{Olspert2015}.
\item 
Dec.2011 is our best SOFIN season with a 60\% phase coverage. The inverted map
shows both cool spots at equatorial and 
high latitudes with middle latitudes devoid of spots. 
The strongest temperature minimum occurs at around phase 0.1, but the high-latitude
spot structure is again very elongated, possibly forming an asymmetric 
cool polar cap on the star.
\item The Dec.2012 season has a phase coverage of 40\% and the single cool spot at mid-latitudes appears 
in the largest gap between observed phases and may thus be an artefact.
The typical high-latitude structures seem not to be present any longer.
\item 
The Nov.2013 map shows again a checkerboard pattern and the phase coverage 
is only 40\%. 
Cool spots in this map appear at mid-latitudes and are paired with warmer spots 
that are probably artefacts.
Again, high-latitude structures are absent.
\end{itemize}

\begin{figure}[h!]
\includegraphics[width=\columnwidth]{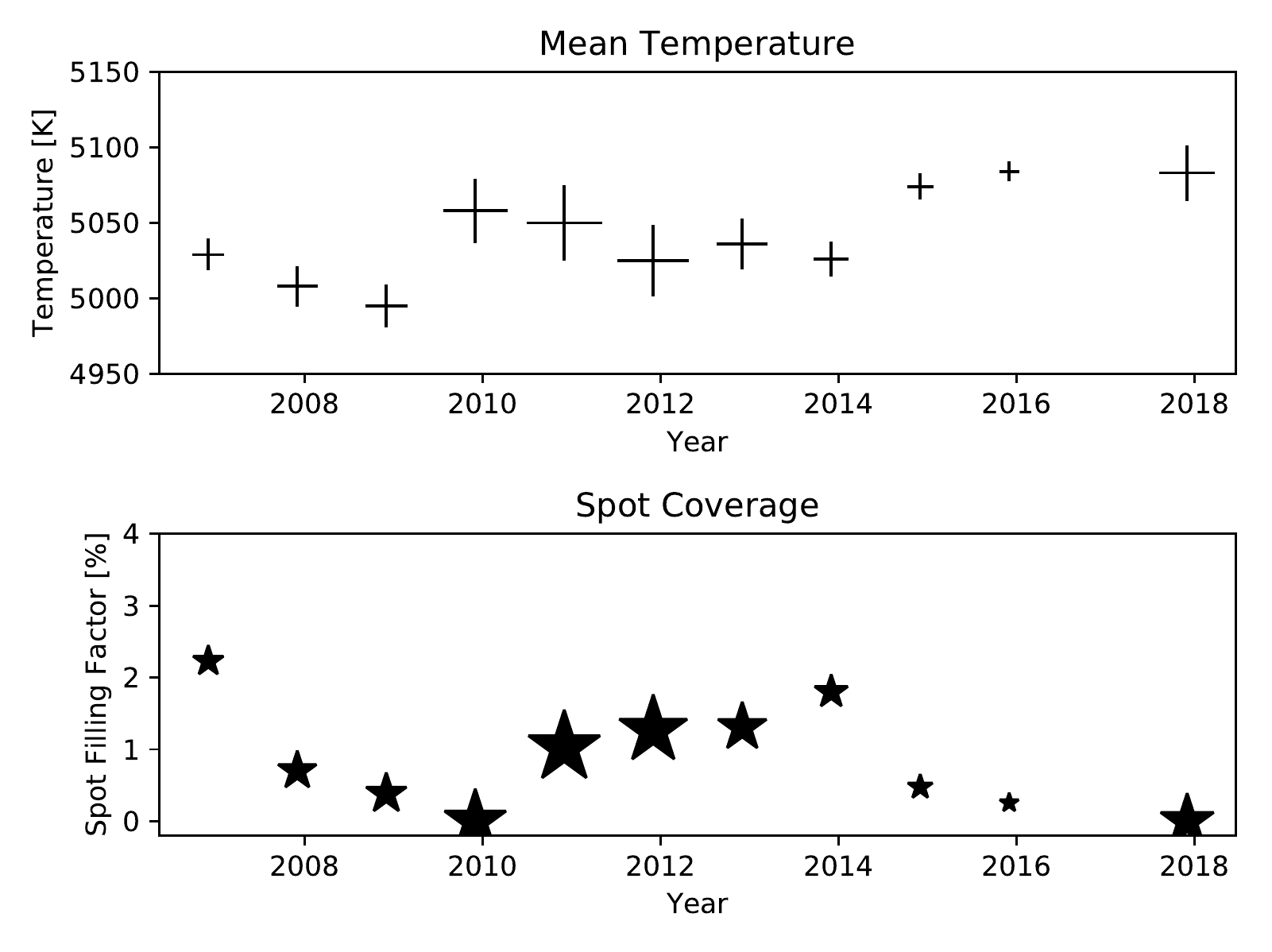}
\caption{Mean temperature (top) and spot filling factor (bottom) of the 
derived surface temperature maps. 
The symbol size is proportional to
$(\langle\textrm{S/N}\rangle \times f_\phi)^{2}$.}
\label{fig:temp}
\end{figure}

\begin{figure}[!h]
\includegraphics[width=\columnwidth]{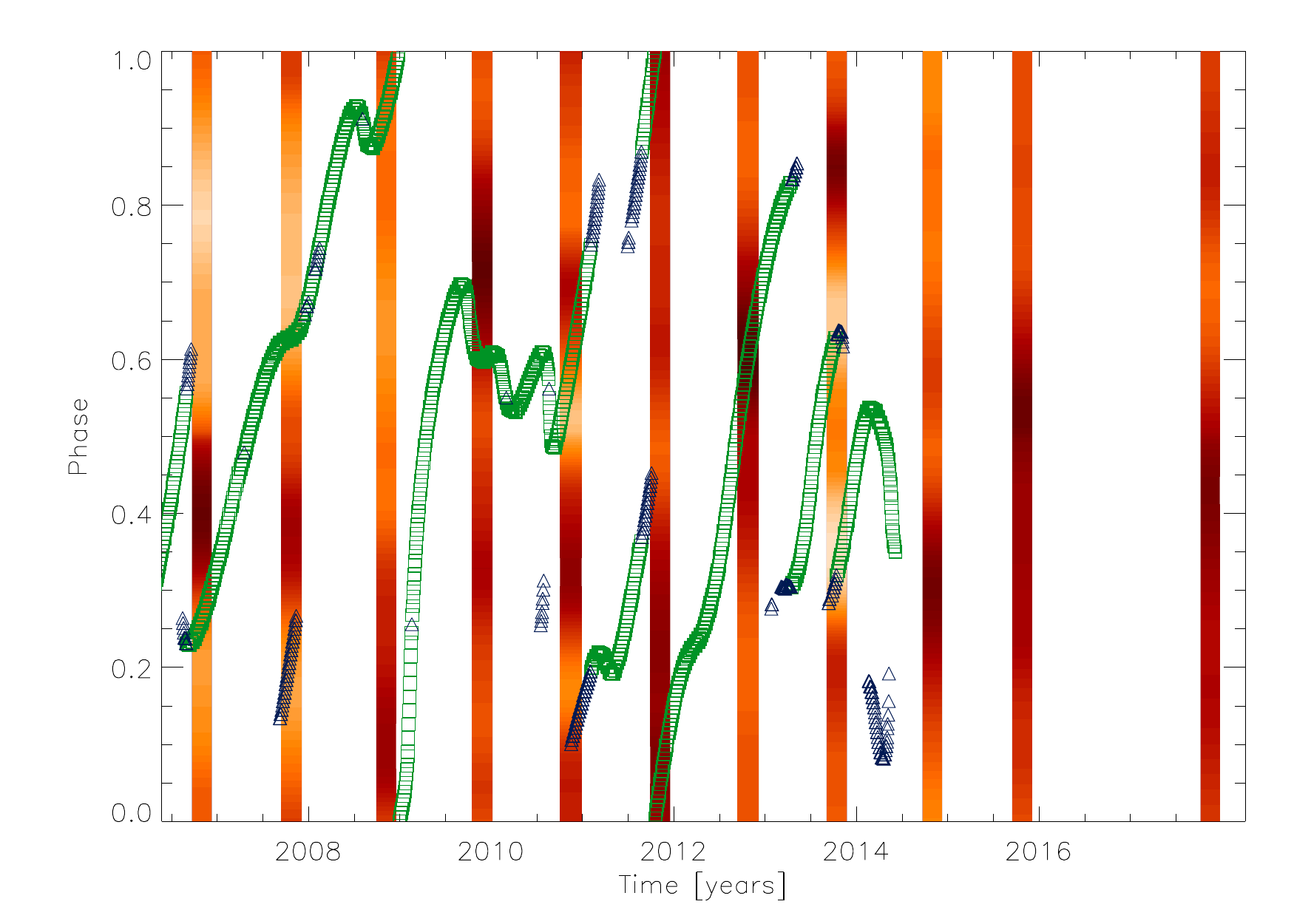}
\includegraphics[width=\columnwidth]{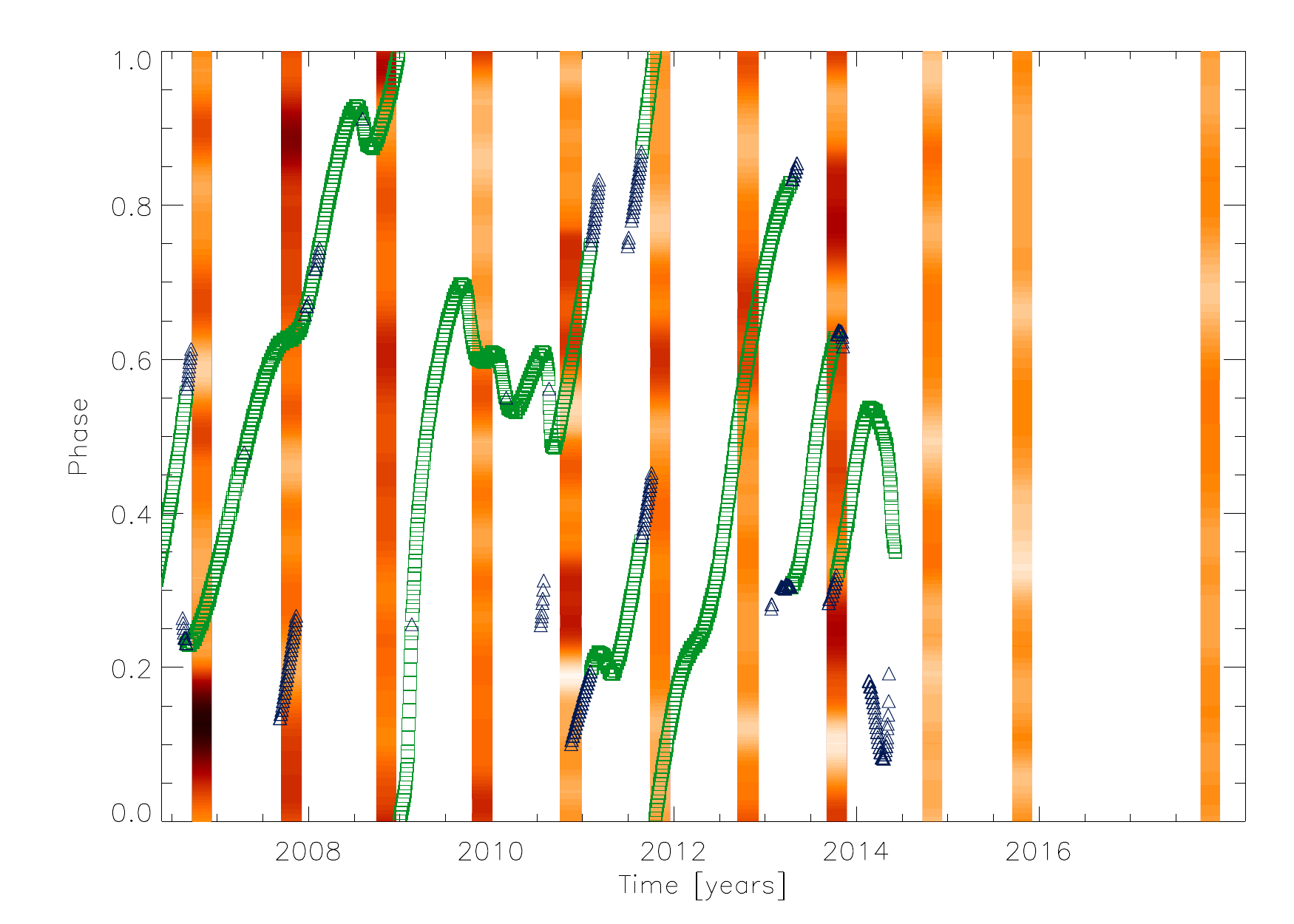}
\includegraphics[width=\columnwidth]{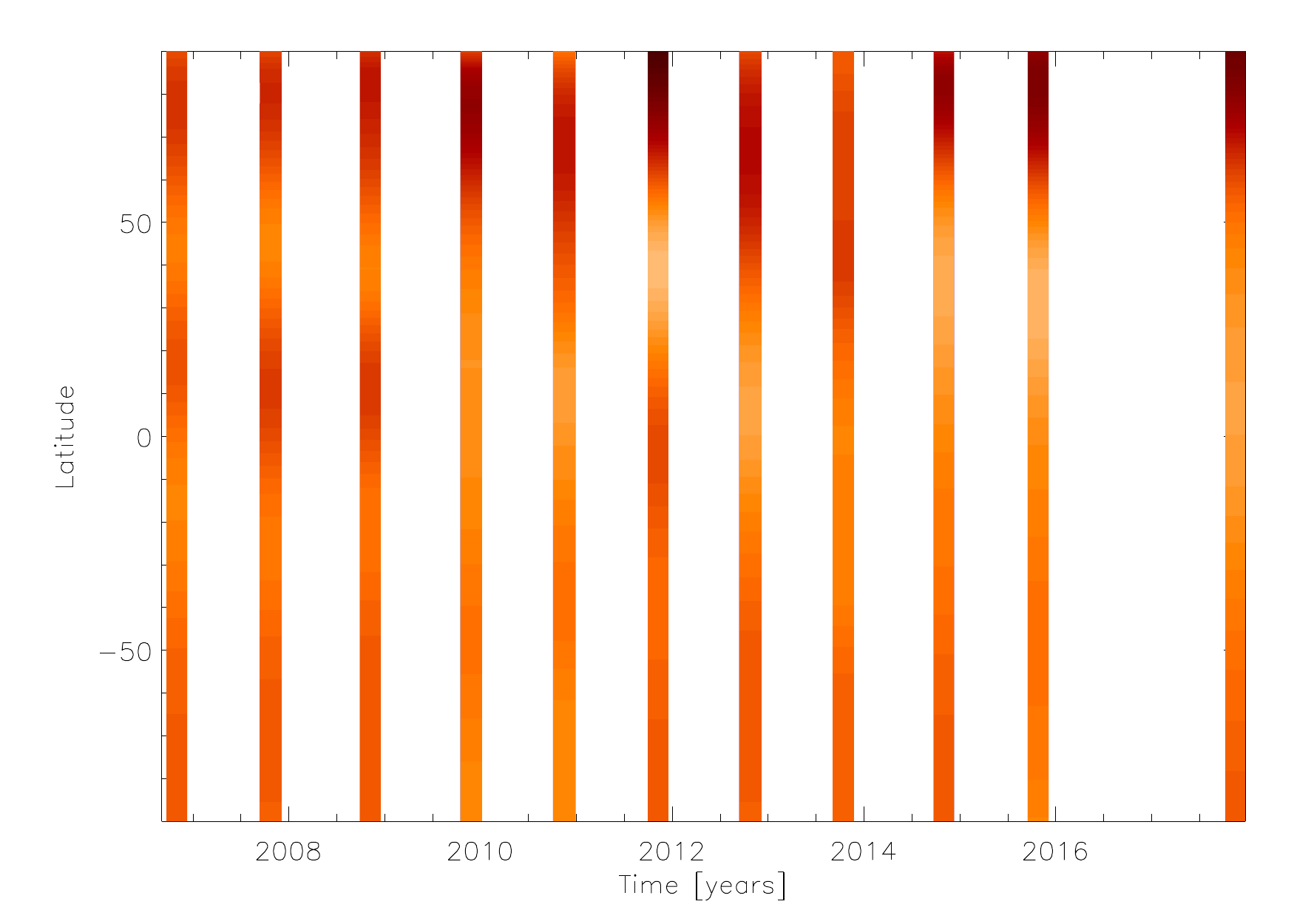}
\caption{Upper two plots show phase-time diagrams computed from the Doppler 
images by averaging them over latitude ranges. 
The temperature range for both is 4527--5148\,K.
In the top panel we have averaged over the range 45--90\degr, 
while in the middle panel 0--45\degr. 
We overplot with the phases of the photometric minima derived by \cite{Olspert2015}, 
the green squares showing the primary minima and the blue triangles the secondary minima.
The bottom panel shows the latitude-time diagrams computed by averaging the Doppler 
images over the whole longitude range. 
The temperature range is 4464--5230\,K.
}
\label{fig:phasetimes}
\end{figure}

\subsection{Temperature maps from FIES observations, 2014, 2015, and 2017}\label{FIES}

Figure \ref{fig:di_set2} shows the FIES maps where three additional \ion{Fe}{I} lines 
were selected to overlap with those used by \cite{cole2014b}. 
\begin{itemize}
\item Phase coverage for Dec.2014 is again poor, with $f_\phi$ of only 39\%.
After several years of absence of high-latitude activity, we now recover an
extended high-latitude spot structure with a temperature minimum around
the phase 0.3. Even though the phase coverage is poor, this structure
coincides with the observed phases and is most likely real. Whether it
would extend even more in phase is, however, unclear due to the large 
phase gap from 0.6--1.1.
An equatorial spot is also retrieved, but as it is paired with a hot spot, 
it could be an artefact.
\item Dec.2015 has a slightly better phase coverage of 50\%, but the observations 
at $\phi = 0.7$ are noisy. 
There is again evidence of a cool spot near the pole.
The temperature minimum of the retrieved structure occurs in a relatively large phase
gap, hence the real longitudinal extent and the exact location of the temperature minimum
remain uncertain.
\item Dec.2017 is considered our best map, with eight observations and a phase 
coverage of 68\%. This map shows a 
cool spot near the polar region, the temperature minimum occurring at around phase 0.4, 
and no spots near the equatorial region. 
\end{itemize}

\subsection{Overall behaviour and comparison to earlier works}\label{Comparison}

Figure \ref{fig:temp} shows the changes in the mean temperature 
and spot filling factor over time.
The symbol
size is proportional to
$(\langle\textrm{S/N}\rangle \times f_\phi)^{2}$
so that larger symbols emphasize the degree of confidence 
arising from higher S/N and better phase coverage. 
The spot filling factor was calculated as the percentage of the 
surface area covered by spots, defined as 
regions colder than 4500\,K.
As previously shown by \cite{Willamo2019} and \cite{Hackman2019}, 
the relative changes in spot coverage are not very sensitive to the 
defined spot temperature.
From the top panel of Figure \ref{fig:temp}, it can be seen that 
there is a trend of increasing mean temperature, which corresponds 
fairly well with the observed brightening of the star between 2006
and 2014 seen in Figure~\ref{fig:photometry}.
Because spot coverage is overestimated with poor phase coverage, only 
the large symbols are reliable 
and thus we can really only conclude that the spot coverage around 
Dec.2011 was greater than the virtually unspotted season of Dec.2017.
Nevertheless, the slight hint of an overall decreasing trend of spot coverage 
is consistent with the increase in 
the mean temperature, and hence supports our
hypothesis of the star entering a low activity state.
Moreover, if we compare our results to the spot coverage results for \cite{cole2014b}, 
we find that LQ~Hya is less spotted overall during 
these more recent observations.  

Spot latitudes, particularly those at lower latitudes, should 
be treated with some skepticism because of the poor phase coverage
for most of the observing seasons and low latitude spots paired with 
hot spots can be treated as possible weakly cool spots or artefacts.
However, we consider the spot phases to be robust.
To try to minimise the effect of the low latitude spot structures,
most likely artefacts, we split our latitudinal averages of the temperature
at each phase into two categories: 
high-latitude spots, those above $45\degr$, 
and low-latitude spots, those in between the equator and $45\degr$.
No spots below $-65\degr$ latitude can be observed due to the 
inclination angle of the star
and so we use the northern hemisphere only for the averages.
The reports of various authors that the mid-latitude region is void of spots 
also motivates this approach
\citep{Strassmeier1993,rice1998,donati1999,donati2003b,cole2014a}.
We then plot these in Figure~\ref{fig:phasetimes}
against the photometric minima of 
\cite{Olspert2015}. 
From  the top and middle panels of 
Figure~\ref{fig:phasetimes}, it can be seen that the photometric minima are in 
better agreement with the phases of high latitude spots (top) than with those at  
low latitudes (middle).

In the bottom panel of Figure \ref{fig:phasetimes}, we take the 
average over each latitude to see what latitudes spots tend to 
form at. 
During all observing seasons except 
Dec.2010, 2012, and 2013, we find that there are cool spots 
at high latitudes (Figures~\ref{fig:di_set1}, \ref{fig:di_set2}).
These observations without high latitude spots 
suffer from poor phase coverage, so we cannot exclude the possibility that 
such spots also exist for these seasons, just that they did not fall near the phases of our observations.
However, the typical high-latitude spot structures have usually large phase extents,
and during other observing seasons with equally poor phase coverage we do recover them.
Hence, it is possible that during the chaotic state beginning in 2010, the high-latitude
spot structure was suppressed. In 2014 and from there onwards, the high-latitude spot structure recovers.
Of the two maps with the best phase coverage, Dec.2011 and Dec.2017, 
we find that Dec.2011 has both spots at high and low latitudes,
while Dec.2017 has only the spot at a high latitude.
We see some bimodality (spots appearing at only high and low 
latitudes) in Figure~\ref{fig:phasetimes}, bottom,
for the SOFIN observations but this is not as pronounced for 
the FIES observations. 
However, this must be taken with a grain of salt because, 
as previously discussed, latitudinal information particularly for 
low latitude spots is lost with poor phase coverage.

\cite{Rice2000} found that phase gaps as large as $100\degr$ still reproduced 
spot phases from an artificial map containing large spots, 
although the spot temperatures and shape were changed. 
Our phase gaps are somewhat larger for some seasons, but we find the areas
between our larger phase gaps to be relatively smooth in temperature gradients
with the exception of the Dec.2008 and Dec.2012 maps, which have cool 
spots in the large phase gaps. 
\cite{Lindborg2014} used a temperature map of DI Piscium from an observing season 
with good phase coverage and removed all but five observations and found that 
a previously weak cool spot increased in contrast and a corresponding 
hot spot formed at the same phase.
Thus we would expect then that our low-latitude, cool spots, if physical, when paired with hot 
spots are actually weaker cool spots at those phases and the hot spots 
are likely not physical. 
In \cite{cole2014b} it was seen that poor phase coverage increases the  
temperature contrast $\Delta T$ by 300\,K which results in an overestimate of spot 
coverage. The mean temperature however changed by only 10\,K. 
Thus we expect our temperature differences to be 
dependent on the phase coverage, increasing with poorer coverage. 
As the spot filling factor is calculated from this quantity, 
this result would be correspondingly shifted to a higher value and 
hence our results are more of an upper limit.
However, the mean temperature was found to shift by very little in all cases,
and so our results of mean temperature are considered more robust than 
the spot filling factor and the spot temperature. 

The bimodality of the spot distribution
with very few spots at mid-latitudes was also observed in DI maps from earlier 
epochs, such as those by \cite{rice1998} and \cite{cole2014b}.
We do not find a band of spots around the $+30\degr$ latitude as was the 
case for \cite{Kovari2004}, although this may be explained by our lower value for 
$v \sin i$. 
\cite{Soriano2017} also performed a Doppler imaging analysis of LQ~Hya 
during December 2011 with some overlap of similar spectral lines for 
\ion{Fe}{I} and \ion{Ca}{I} as in our study. 
The Doppler maps closest in time to our Dec.2011 map have two near-polar spots 
and one spot close to the equator. 
Our maps seem to agree with this spot configuration, with our Dec.2011 map showing a 
large spot near the polar region and another 
spot closer to the equator.
Their phase for the low-latitude spot is different from our 
phase by $\sim 0.5$ when accounting for the different ephermeris, but 
the midpoint of the two high-latitude spots does 
match the phase of our elongated high-latitude feature. 
This map also had decent phase coverage, so the equatorial spot is likely physical. 
A persistent spot near the pole is also observed in the ZDI maps of 
\cite{donati1999} and \cite{donati2003}, 
and this was found to be similarly nonaxisymmetric. 

Contemporaneous observations in photometry 
to measure periodicities and study the phenomena of active longitudes and flip-flops 
are of interest when examining the maps. 
\cite{Olspert2015} found evidence of a flip-flop during late 2010, and phase disruptions 
during 2012 and 2013. They also found evidence of a phase drift during 2008.
\cite{lehtinen2012} found flip-flop behaviour during their late-2010/early-2011 
observations.
The recovered spots in our Dec.2010 map are close to the phases of the light-curve minima 
and so while the higher temperature contrast is likely an artifact, the spots 
themselves would be physical, revealing a rather chaotic surface during this epoch 
corresponding with the chaos from flip-flops and phase jumps found in the photometry. 
Visual inspection of the spectral lines for Dec.2010 in Figure~\ref{fig:di_set1}
reveals significantly changed line profiles in between rather small changes in phase, 
supporting the chaotic surface temperature map of this season. 

The photometric minima, indicating active regions, 
have been well-matched 
to indicators of chromospheric activity for LQ~Hya. 
\cite{cao2014} matched their phases of observations from 2006--2012 
of plage regions to the photometry of 
\cite{lehtinen2012} and found an increase in chromospheric activity corresponded 
to a decrease in photometric magnitude. 
\cite{cao2014} also found that the chromospheric activity level in general steadily 
decreased throughout their observations from 2006--2012, 
which matches the increase in magnitude in 
photometry seen during this time as well as the increase in the mean temperature in 
our maps.
\cite{cao2014} found a plage region in Febuary 2012.
If we convert their  observations to our ephemeris in Equation~\ref{ephemeris}, 
we find their plage region occurs at $\phi= 0.26$ which coincides
with our low-latitude spot near that phase. 

The zonal models of \cite{Livshits2003} limit spots to two latitudinal belts 
symmetric about the equator and examines the shift in the upper and lower 
latitudinal boundaries, a rough approximation of the butterfly diagrams of the Sun.
For the epochs 1983--2001, they find that a rise in activity level corresponds to 
an equatorward drift of the lower latitudinal boundary of the spot zones, where the relative 
spotted area is used as the indicator of activity. 
The upper boundary
of the zone remained somewhat constant and was relatively low ($< 50\degr$). 
However, we find mainly high-latitude or low-latitude spots and not many in the 
$35\degr$--$50\degr$ range, and from Figure~\ref{fig:phasetimes}, bottom, we see that 
our low-latitude spots get weaker from Dec.2014 to Dec.2017
while the high-latitude spot remains. 
During the 
chaotic period between Dec.2010--2013, spots and both higher and lower latitudes 
appear and disappear from season to season, although some of this may be due to poor 
phase coverage.

\section{Discussion on the data quality and effects on the maps}\label{section:discussion}

Poor phase coverage is the primary source of artefacts in most of our maps. 
Spots located at or near the observed phases, indicated by vertical dashed lines in 
Figures \ref{fig:di_set1} and \ref{fig:di_set2}, 
are likely physical as the S/N in these observations were high. 
Furthermore, the evidence of spots is supported by visual
inspection of the spectral lines in each figure. 
For phases not adequately covered by observations, the inversion
program lacks information, 
and the result in cases where spots are near the observed phases 
is the checkboard pattern seen 
in, for example, the Dec.2010 map (Figure \ref{fig:di_set1}).
The inversion program does not introduce spurious spots 
into areas between the phase gaps but it may increase the temperature 
contrast of spots further away from the observed phases, such as those seen in our 
Dec.2008 and Dec.2012 maps (Figure~\ref{fig:di_set1}).

When phase coverage is poor, the recovered latitudes of spots 
are also less precise, particularly for spots at lower latitudes. 
Additionally, the ability of the inversion program to distinguish between low latitude spots above 
or below the equator worsens with poor phase coverage.
Furthermore, LQ~Hya is an active star, and while spot 
structures generally persist for a month or more, 
sudden changes are possible, as is indicative from photometry especially for 2010-2013.
Therefore, due to such rapid changes, spots at higher latitudes 
may be interpreted as spots at lower latitudes. 
We consider the effects of rapid variability by examining our Dec.2007 and 
Dec.2017 maps, both of which had two observations close in phase but distant in 
time by eight and three days, respectively.
Running the inversion excluding first the later observation and then the 
earlier observation showed no significant difference in the resulting temperature 
maps around the observed phases beyond a slight change in spot shape. 
We also keep in mind that the star seems to be decreasing in activity level,
and all our observing seasons are 12 days or less, making this 
source of artefacts less likely. 
Because of the limitations of the inversion method and poor phase coverage,
the spot phases and spots at high latitudes should be more reliable 
than low latitude spots.

\section{Conclusions} \label{sec:CONCLUSIONS}

We have calculated surface temperature maps for LQ~Hya for 11 observing seasons
ranging from December 2006 to December 2017 epoch using the DI technique.
We summarize our findings as follows:
\begin{itemize}
\item Seasons with poorer phase coverage are less reliable, particularly 
for quantities like the
spot latitude and especially for the spots seen at near-equatorial regions. 
However, spot phases are still robust, and the high-latitude spots 
are likely physical, while the accuracy of 
low-latitude spots depends on the phase coverage.
\item We find an increase in mean temperature throughout the observing 
seasons with only a slight dip between 2009 and 2011. 
This matches the increase in the observed magnitude of the 
star during this time, indicating a decrease in stellar activity. 
\item The primary and secondary minima from concurrent photometry 
better match the phases of high-latitude spots than the phases of 
low-latitude spots. 
\item There appears to be a bimodal spot distribution over latitude, which is 
in agreement with previous DI and ZDI maps of LQ~Hya. 
However, the lower latitude spots become weaker as the activity 
level of the star decreases while the higher latitude spots persist
during the observing seasons with FIES. 
Both higher and lower latitude spots 
appear and disappear from season to season with SOFIN.
\item 
Photometry indicates an especially chaotic epoch of spot evolution during
2010--2013, with rapid spot migration in the rotational frame, and abrupt phase changes, that
were characterised as flip-flop events by \cite{Olspert2015}. Spectroscopy during
this time indicates strong line profile variability, indicative of large
spottedness, but the maps show checkerboard patterns. 
Also, the high-latitude spot structures disappear
for 2010, 2012, and 2013. Characterising the spot structures from Doppler
images is very challenging during this epoch.
\end{itemize}
Based on our results, 
LQ~Hya seems to be
approaching an activity minimum. 
During a higher activity state, investigated through Doppler imaging by \cite{cole2014b}, 
temperature maps showed mean temperatures
$\sim200$\,K lower than in this study. 
Those temperature maps, while having a lower S/N on average, show more 
disruption and bumps in the spectral lines than the ones presented here.
Even though the activity is decreasing, the spot structures are still largely chaotic 
where the lower latitude spots do not appear to form at any preferred longitude. 
This matches what is found in numerical simulations, such as those by \cite{Viviani2018}, 
where rapid rotation led to a dominance of the non-axisymmetric portion of the magnetic field.
Their solutions with roughly twenty times solar rotation rate showed strong high-latitude 
magnetic fields organised in two active longitudes of opposite polarity on the same 
hemisphere, while a weaker near-equator activity belt was accompanying these structures.

\begin{acknowledgements}
EC acknowledges funding from the the Deutsche Forschungs-Gemeinschaft (DFG project 4535/1-1). 
MJK and JL acknowledge the Academy of Finland "ReSoLVE" Centre of 
Excellence (project number 307411) and the Max Planck Research Group "SOLSTAR" funding.
OK acknowledges support by the Knut and Alice Wallenberg Foundation 
(project grant “The New Milky Way”), the Swedish Research Council (project 621-2014-5720), 
and the Swedish National Space Board (projects 185/14, 137/17).
\end{acknowledgements}

\bibliographystyle{aa}
\bibliography{lqhya.bib}
\end{document}